# Assessing a human mediated current awareness service

*Zeljko Carevic*[1], *Thomas Krichel*[2], *Philipp Mayr*[1]

[1]GESIS – Leibniz Institute for the Social Sciences
Unter Sachsenhausen 6-8
50667 Köln, Germany
firstname.lastname@gesis.org

[2]Open Library Society
34-20 78th Street #3D
Jackson Heights, NY 11372-2572
krichel@openlib.org

**Abstract**

In this paper, we present an approach for analyzing the behaviour of editors in the large current awareness service "NEP: New Economics Papers". We processed data from more than 38,000 issues derived from 90 different NEP reports over the past ten years. The aim of our analysis was to gain an inside to the editor behaviour when creating an issue and to look for factors that influence the success of a report. In our study, we looked at the following features: average editing time, the average number of papers in an issue and the editor effort measured on presorted issues as relative search length (RSL). We found an average issue size of 12.4 documents per issue. The average editing time is rather low with 14.5 minutes. We conclude that the success of a report is mainly driven by its topic and the number of subscribers, as well as proactive action by the editor to promote the report in her community.

# 1. Introduction

According to Bates (1996) selective dissemination of information (SDI) "is a technique whereby bibliographic citations or copies of new materials received in the library are selectively sent to individual researchers. The selection is based on profiles prepared of each researcher's interests." SDI is a technical procedure which goes back to the early 1970s (see e.g. Leggate, 1975) when the first electronic databases like MEDLARS have been established. These early systems already provided rudimentary alerting services which could be printed out on hardcopy in the library. The effectiveness of these SDI services has been measured from the very beginning (e.g. Packer and Soergel, 1979). The updated term for SDI is now current awareness or alerting service, however, the basic functionality remained the same. Users still want to be alerted regularly when new relevant information on their interest arrives. Interestingly, these services are still heavily used (e.g. Google Alert) and generate a value-added for researchers (Hienert et al., to appear).

The success of these services motivated us to analyze the data in the intellectual editing process of a large current awareness service, namely the RePEc digital library. The main concern of this paper is the basic description of editor behavior and the application of different measures to explore the paper selection pattern of different editors. In the empirical part, we calculated the average editing time and interrelate the amount of subscribers with issue size and editing time. In addition, we started to apply different measures to describe the relevance distribution in these alerting services.

# 2. RePEc and NEP

## 2.1 RePEc

The RePEc digital library, see http://repec.org, is an aggregation of data held by over 1,700 contributing "archives". These archives provide metadata for working papers and journal articles in economics. In fact, as pointed out by Krichel (1997), the creation of a current awareness service was his major motivation for working on what was to become NeEc, a precursor of RePEc.

Nowadays, RePEc archives also provide other data, but that other data is not of concern for this paper. The document metadata for working papers usually contains links to the full texts.

## 2.2 NEP: New Economics Papers

"NEP: New Economics Papers" is a creation of Thomas Krichel. It is dating back to 1998. The idea of NEP is to take new additions to the working paper stock of RePEc and filter them into subject-specific reports. To understand how it works, it is important to distinguish between issues and reports. New additions to the RePEc working paper stock form an issue. The issue takes the name of the date on which it was composed. It usually contains all the new additions to RePEc between the last issue and the current date. There are exceptions, of course. Sometimes an archive adds old papers. Papers that do not have a date in the metadata may be excluded. Whether they are or not, depends on the whim of the person overseeing the process. That person carries the title "general editor". The result of the general editor's work is an issue of a report that contains all recent new working papers in RePEc. This is what NEP calls a nep-all issue. It is not available for subscription at this time, even though in the early days of NEP, it was. A subject report issue is initially essentially a copy of a nep-all issue. Just the name of the report, the name of the editor, etc. are changed. The set of papers that it proposes to an editor are the same as in the nep-all issue. An editor carries out two operations on the report issue. The first, required step is to remove papers from the nep-all issue that do not fit into the subject of the report. The second, optional step is to sort the remaining paper in a way that may bring the most interesting paper to the top.

The oldest known nep-all issue is from May 4, 1998. It contains 13 papers. Since then nep-all issues have appeared roughly on a weekly bases. Subject reports have appeared over time as volunteer editors have become available. Still there has been a concern of the NEP administration to cover economics as a whole. Krichel and Bakkalbasi (2005) provide a statistical analysis with this aim in mind. At the time of writing close to 90 reports are in business. Bátiz-Lazo and Krichel (2012) have more on the history of NEP. The NEP web site is at http://nep.repec.org.

## 2.3. Ernad

In 2004, Thomas Krichel hired Roman D. Shapiro to produce a purposed-built system to create NEP report issues. The resulting system is called “ernad”, editing reports on new academic documents. There were two motivations behind ernad. First, Krichel felt that a better monitoring system was required to understand how the system was working. It would allow us to write papers about how it works. It would also be useful to monitor editor performance, to remedy some of the issues in the review of NEP by Chu and Krichel (2003). Second, Krichel forecasted that with the growth of RePEc, there would come a moment where examining all the papers would be too time-consuming for volunteers to do. Thus, the introduction of statistical learning to aid the human effort would be required. Roughly speaking, ernad is looking at past issues of a report to produce an order of papers, ranked by the likelihood that they will be appearing in the current report issue. If presorting works well, the editor can pay a lot of attention to the top of the presorted issue and spent little to no time on the papers at the end. Without presorting, the composition of a subject report issue would take a long time because the editor has to give a minimum of attention to each document. Obviously, when NEP starts a new report, no learning data is available. The rookie editor of a new report will have to wade through about a 1,000 papers in the initial issue without any help.

# 3. Materials and Methods

Table 1 shows some statistics of the NEP service. Currently, NEP consists of over 90 active reports with 75,239 subriptions. In total we analyzed 38,719 issues in 90 reports which have been published from December 2005 until November 2014. An issue contains an average of 12.4 papers although this number varies depending on the topic. The macroeconomics report (nep-mac) for instance contains about 50 papers per issue whereas the report for sports and economics (nep-spo) only contains an average of 2.5 papers. 90% of the issues appear to be based on presorted data. The remaining 10% are based on unsorted nep-all issues.

Table 1: NEP statistics

| #Reports | #Subriptions | Avg. #Subriptions | Avg. nep-all size | Avg. Issue size | Reports pref. sort |
|---|---|---|---|---|---|
| 90 | 75239 | 836 | 488 | 12.4 | 90% |

## 3.1 Editing stages

After login in, ernad shows the editor a list of issues to work on. Each issue, as represented by a date, comes with three buttons. These are labeled "presorted", "unsorted" and "delete", respectively. When selecting "presorted" the editor is given a list of all the papers in the corresponding nep-all issue as presorted by the system. When selection "unsorted" the editor is given a list of all the papers in the corresponding nep-all issue in the same order as they were in the nep-all issue. The editor has to select each paper individually by clicking its check box. No paper is selected by default. If the editor selects no paper for inclusion, ernad does not proceed to the next stage. Instead, the paper selection screen features a back button. There, the editor can go back to the issue selection screen to delete the issue. After selecting papers the editor reaches the ordering stage. Here, the editor can change the order of papers in the report issue, or delete papers in the issue. The final stage is the sent report issue. The ernad system saves copies of the report issue files in each of the four editing stages (see Figure 1). Each stage is a directory in the file system. Since every stage may be repeated, several versions of a report issue file for the same issue may be present in the directory. In this case, for our results, we use the latest one.

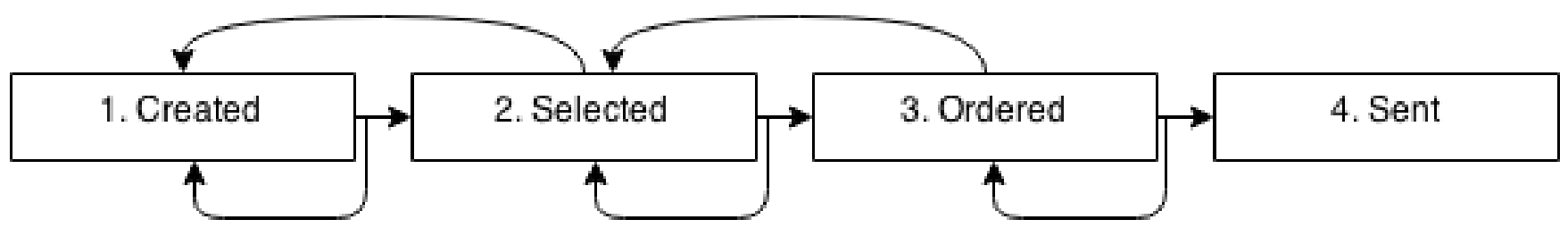

Figure 1: Editing stages in NEP issue creation

## 3.2 Preselection of report issues

Editors can start selecting relevant papers for their issue but do not necessarily need to work on it continuously. In other words, an editor could start creating an issue but interrupt her work for some reason and continue working on it a few days later. This feature of ernad is very convenient for the editors. But it makes it hard for us when we want to find out how long an editor has been working on an issue. Looking at the editing stages in NEP we cannot distinguish between issues which were created without interruption and issues where the editor took a long break. Taking issues that were interrupted during their editing into account would distort the results. We therefore define a threshold for the maximum editing duration. We consider every issue created below this threshold as continuously and therefore valid. To estimate our threshold we split the editing duration of all issues into chunks of 3 minutes. Figure 2 shows that about 88.9% of the total 38,719 issues are created in less than 90 minutes. We consider the issues generated in more than 90 minutes as interrupted and therefore exclude them from the further analysis.

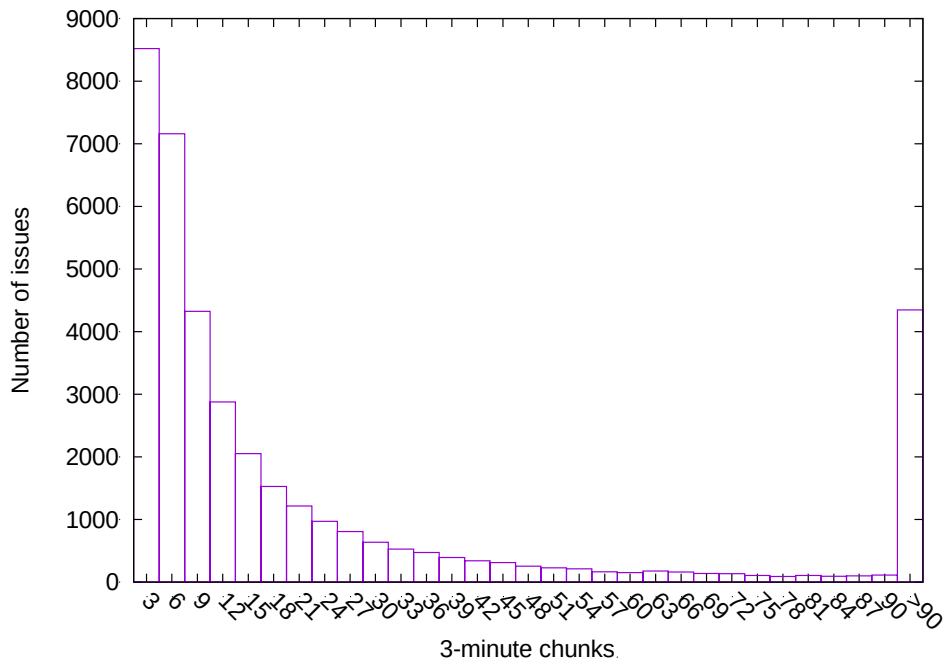

Figure 2: Issues separated into 3-minute chunks

# 4. Results

In the following we try to gain an insight into the process of creating an issue. Therefore, we created four types of NEP-specific measures. First, we looked at the *average editing time* of editors when creating an issue. Second, we build *ratios of subscribers* with *issue size* and *editing time*. In this way, we want to find out if these ratios influence the success of a report. The third and fourth measure aim to describe the efficiency of presorting and the effort an editor invests in creating an issue. Therefore, we applied the *precision at n* (P@n) and the new *relative search length* (RSL) measure for each issue.

## 4.1 Average editing time per report

We now describe the average editing time for creating an issue. Figure 3 shows the distribution of the average editing time per issue for all 90 NEP reports. We found that editors invest on average 14.5 minutes per issue, with a standard deviation of 10.1 minutes. A maximum of 53 minutes per issue was found for report nep-ets (Economic Time Series) and a minimum of 2.5 minutes for report nep-res (Resource Economics).

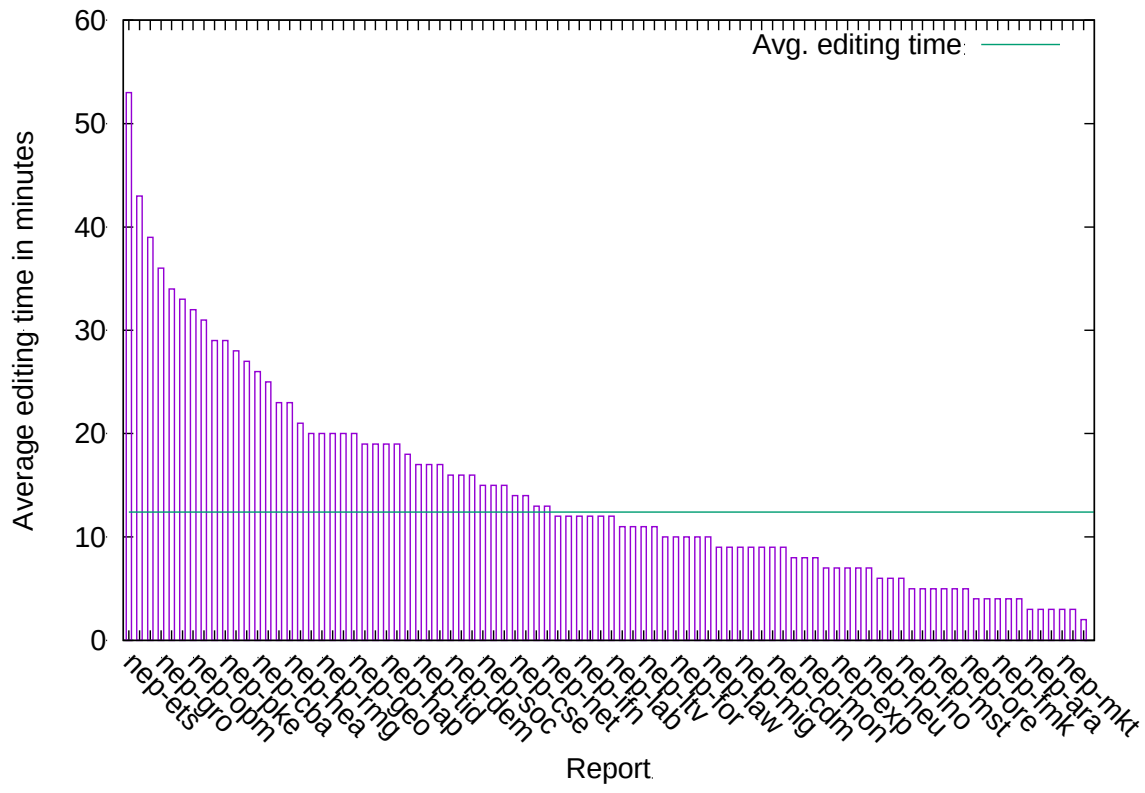

Figure 3: Average editing time per report

## 4.2 Ratio of subscribers with issue size and editing time

The popularity of a report can be measured e.g. by the number of subscribers. When looking at the number of subscribers a huge scattering between the reports can be found. The most successful report is nep-his (Business, Economics and Financial History) with approx. 7,000 subscriber. The lowest number is for nep-cis (Confederation of Independent States). It attracts only 78 subscribers. The number of subscribers is dependant on the topic, the age of a report and of course the editor. We assume that reports that exist longer had more time to gather subscribers. In the next step we want to investigate if there are any other criteria that influence the success of a report. To this end we look at two factors, the editing time and the number of papers in an issue, i.e. the issue size.

In Figure 4a we compared the average editing time with the number of subscribers. Our hypothesis is: the more time an editor invests in creating an issue, the more value-added the report contains, so it should attract more subscribers. Figure 4a shows that there is no correlation between the editing time and the number of subscribers (Pearson correlation coefficient: 0.01). For example one of the most popular reports nep-edu (Education) shows a rather low editing duration of 4.5 minutes. A second idea is that the more papers an issue contains the more subscribers the report obtains. This seems logical as a broad topic, with a lot of papers will be more interesting to a wider range of subscribers. As illustrated in Figure 4b, again there is just a poor correlation between these factors (Pearson correlation coefficient: 0.21).

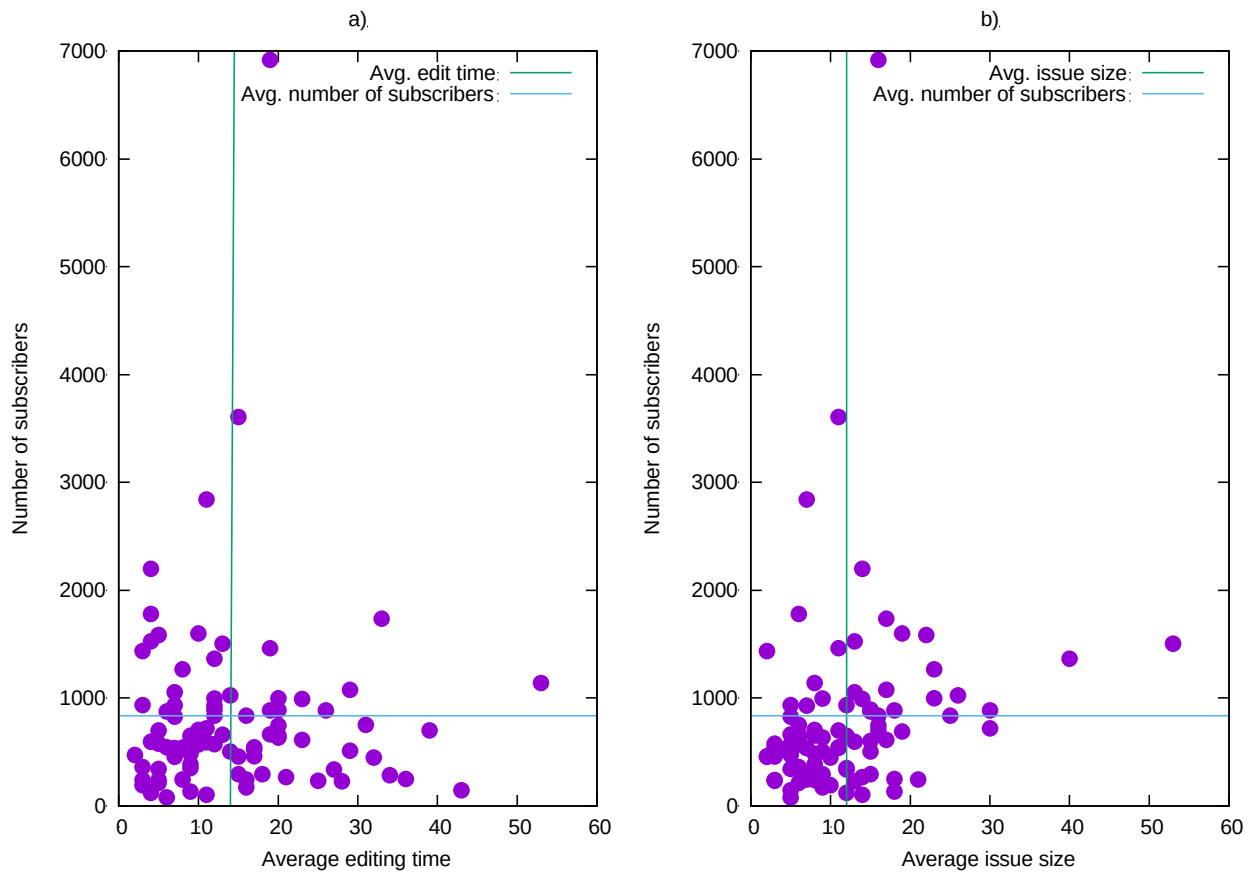

Figure 4: a) Ratio of subscribers with editing time, b) Ratio of subscribers with issue size.

**We summarise:** the average issue size and the average editing time of a report have no influence on the popularity of a report.

## 4.3 Relevance distribution in presorted nep-all

In a next step we want to measure how much effort is invested in the manual selection and sorting of papers during the issue creation by the editors. We are looking at the differences between the *presorted* nep-all and the final distributed issue (see sent in Figure 1). We use the common precision at N (P@N) measure which is described in equation 1 where N is set successively to document 5, 10, 15 and 20. A document is considered relevant if its index position in nep-all is <= N. The following example is used to illustrate the calculation of P@N. The editor receives a presorted nep-all, which comprises a set of documents. From this list the editor now chooses a ranked set of relevant documents M={(D1, 4), (D2, 1), (D3, 7), (D4, 3), (D5, 9)} where each item is a tuple containing the document and the index position in nep-all. In this example D1, D2 and D4 are considered relevant when calculating P@5 due to its index position being <= N. If an issue contains less than N papers we exclude this issue from the calculation.

$$P@N = \frac{\text{number of relevant document}}{N}$$

Equation 1: Precision at N

The equation for measuring the average P@N (AP@N) for each report and each included issue is described in equation 2.

$$AP@N = \frac{1}{|Q|} \sum_{j=0}^{|Q|} \frac{1}{|R_j|} \sum_{k=0}^{|R_j|} P@N(I_k)$$

Equation 2: Average Precision at N

We accumulate the precision at N for each Issue $I_K$ and build the average by dividing it by the set of Issues in Report $R_J$. The overall average precision is then calculated by dividing the cumulated average precision at N for each Report $R_J$ by the set of valid reports Q. In most reports we find a mixture of issues created using presorted and unsorted nep-all. Therefore, we limit our analysis to those reports that use the presorted algorithm at least 50 times. The AP results for all reports using presorting can be seen in Table 2.

Table 2: Precision at 5, 10, 15 and 20

| **Avg. P@5 (82 reports)** | **Avg. P@10 (64 reports)** | **Avg. P@15 (50 reports)** | **Avg. P@20 (31 reports)** |
|---|---|---|---|
| 0.77 | 0.80 | 0.80 | 0.82 |

It can be seen that the precision values increase slightly with higher N values. Considering the large number of papers in the all-report (see Tab. 1) the precision values are promising and show a good acceptance of the presorted papers. For AP@5 we find a precision of 0.77 what means that about 3 of the top 5 documents are considered relevant. We reached the highest precision values for report nep-env (Environmental Economics) with an AP@5 of 0.99 and the lowest AP@5 for nep-cba (Central Bank) with 0.35. For the top five documents (P@5) 82 of 90 reports were considered valid whereas an increasing N leads to less valid reports (e.g. 31 for P@20). This was expected because only a small amount of issues contain at least 20 papers.

**We summarise:** Editors work comfortably with the presorting in nep-all. The number of papers per issue has no significant influence for the precision.

## 4.4 Relative search length

We now investigate how much effort is invested in the selection and sorting of papers during issue creation. In the following we look at the differences between the presorted nep-all and the final issue. To this end we define the relative search length (RSL). This is the ratio between the highest index position of the last relevant document in nep-all and the length of nep-all. For example, the editor is given an all-report, which comprises 300 documents in presorted order. She now chooses a ranked set of relevant documents M={(D1, 4), (D2, 10), (D3, 7)} where each item is a tuple containing the document and the index position in nep-all. In this example the highest index position of the last relevant document (hin) is 10. Having the hin value at hand we assume that the editor has at least inspected each paper with an index-position less or equal hin. Thus, the relative search length is 10/300 = 0.033.

$$RSL = \frac{hin}{\text{nep-all length}}$$

Equation 3: Relative search length

The equation for measuring the average relative search length for each report and each included issue is shown in equation 4. We limit our analysis to those reports that use the presorted algorithm at least 50 times.

$$Avg.RSL = \frac{1}{|R|}\sum_{i=0}^{|R|} RSL_i$$

Equation 4: Average relative search length

In Figure 5 the average relative search length per report is illustrated. The average rsl score for all analyzed 82 reports is 0.08. The maximum rsl (0.35) was found for the macroeconomics report (nep-mac) which was also the report with the highest number of papers per issue (50). The lowest rsl (0.01) was found for sports and economics (nep-spo) and market microstructure (nep-mst). Regarding the number of papers per issue we can observe that both contain a rather low number of papers per issue (2.5 for the former and 3.9 for the latter). When we correlate the average relative search length with the number of papers per issue we get a relatively strong correlation of 0.61.

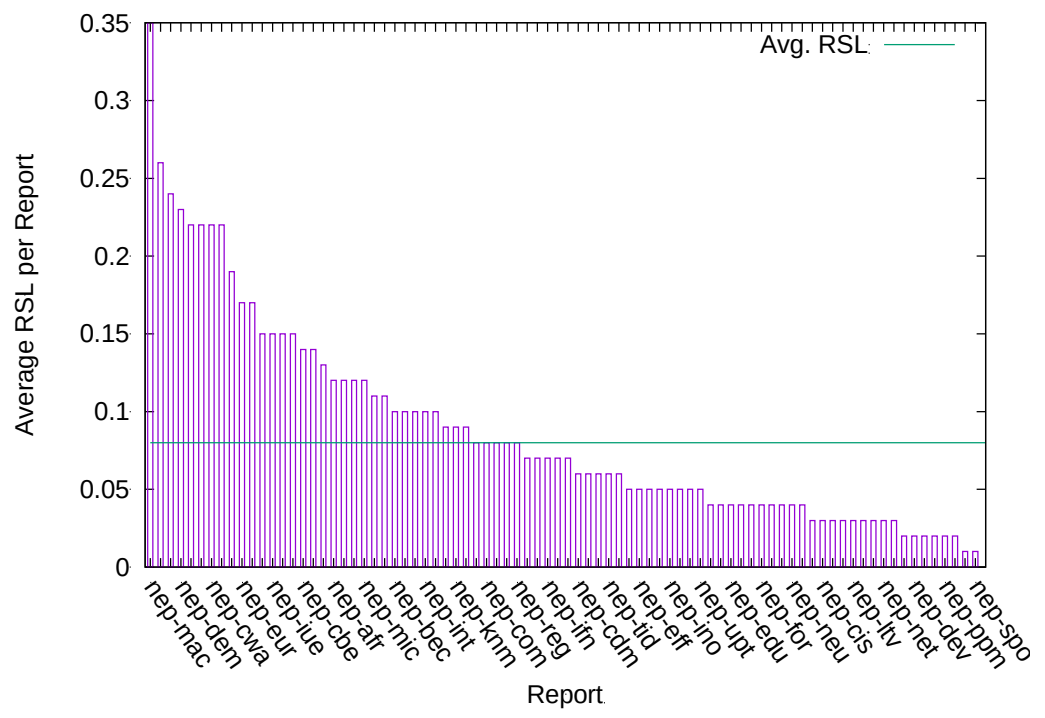

Figure 5: Average relative search length

**We summarise:** The relative search length is comparable low with 0.08 which means that editors select papers from the very upper part of nep-all. Deeper parts of nep-all are usually not inspected anymore.

## 5. Conclusion

In this work we focused on observable system features that do not require any user involvement. To gain a complete view on the editing behaviour of the volunteer editors a more user-centred observation is necessary. One approach would be to use mouse movement observations to see if the editing process is uninterrupted. Furthermore, it would be interesting to investigate why and under what conditions editors consider a paper as relevant. Think aloud observations would help with this. The NEP system generates a lot of data that is available on request. There is data on download request by report and issue available. This data would allow a more precise measure of report success than just subscriber numbers. Subscriber data and papers data could be combined to build a two-level network where nodes are reports and edges are co-subscribers and co-occurrence of papers. It is also possible to look at more sophisticated measures of presorting performance as proposed by Krichel (2007). Thus, NEP provides many opportunities for further research on data that is relatively easily available.

## References

Bates, Marcia J. (1996). "Learning About the Information Seeking of Interdisciplinary Scholars and Students". Library Trends, 45, 155-164.

Bátiz-Lazo, Bernardo, and Thomas Krichel (2012). "A brief business history of an on-line distribution system for academic research called NEP, 1998-2010", Journal of Management History, vol. 18, no. 4, pp. 445-468 March

Chu, Heting and Thomas Krichel (2003) "NEP Current Awareness Service of the RePEc Digital Library" Digital Libraries Magazine, vol. 9, no. 12

Hienert, D., Frank Sawitzki, and Philipp Mayr, (to appear). Digital Library Research in Action – Supporting Information Retrieval in Sowiport. D-Lib Magazine.

Krichel, Thomas (1997). "About NetEc, with special reference to WoPEc", Computers in Higher Education Economics Review, vol. 11, no 1, pp. 19-24.

Krichel, Thomas and Nisa Bakkalbasi (2005). "Developing a predictive model of editor selectivity in a current awareness service of a large digital library", Library and Information Science Research, vol. 27, no. 4, pp. 240-252

Krichel, Thomas (2007). "Information retrieval performance measures for a current awareness report composition aid", Information Processing and Management, vol. 43, pp. 1030-1043

Leggate, Peter (1975). "Computer-based Current Awareness Services". Journal of Documentation, vol. 31, no. 2, 93-115. Retrieved from http://dx.doi.org/10.1108/eb026596

Packer, Katherine H., and Dagobert Soergel (1979). "The importance of SDI for current awareness in fields with severe scatter of information". Journal of the American Society for Information Science, 30(3), 125-135. doi:10.1002/asi.4630300303